\documentclass[runningheads,a4paper,table]{llncs}
\usepackage{multirow}
\usepackage{amssymb}
\usepackage{graphicx}
\usepackage{url}
\usepackage{longtable}
\usepackage{tikz}
\usetikzlibrary{mindmap,trees}
\usepackage{lipsum,adjustbox}
\usepackage{booktabs}
\usepackage{todonotes}
\usepackage{rotating}
\usepackage{lscape}
\usepackage{caption}

\usetikzlibrary{calc,positioning}
\definecolor{rosso}{RGB}{220,57,18}
\definecolor{giallo}{RGB}{255,153,0}
\definecolor{blu}{RGB}{102,140,217}
\definecolor{blu2}{RGB}{0, 137, 204}
\definecolor{blu3}{RGB}{17, 102, 139}
\definecolor{verde}{RGB}{16,150,24}
\definecolor{verde2}{RGB}{117, 200, 16}
\definecolor{viola}{RGB}{153,0,153}
\definecolor{viola2}{RGB}{83,40,213}
\definecolor{myellow}{RGB}{255,233,114}

\newcommand{\keywords}[1]{\par\addvspace\baselineskip
\noindent\keywordname\enspace\ignorespaces#1}

\begin{document}
\mainmatter  

\title{SoK: Blockchain Solutions for Forensics}

\titlerunning{SoK: Blockchain Solutions for Forensics}

%
%
\author{Thomas K. Dasaklis\inst{1} \and Fran Casino\inst{1} \and Constantinos Patsakis\inst{1,2}}
\authorrunning{Dasaklis et al.}

\institute{Department of Informatics, University of Piraeus, Greece\\
    \email{\{dasaklis,francasino,kpatsak\}@unipi.gr}
\and
Athena Research Center}
%
%

\toctitle{SoK: Blockchain Solutions for Forensics}
\maketitle

\begin{abstract}

As the digitization of information-intensive processes gains momentum in nowadays, the concern is growing about how to deal with the ever-growing problem of cybercrime. To this end, law enforcement officials and security firms use sophisticated digital forensics techniques for analyzing and investigating cybercrimes. However, multi-jurisdictional mandates, interoperability issues, the massive amount of evidence gathered (multimedia, text etc.) and multiple stakeholders involved (law enforcement agencies, security firms etc.) are just a few among the various challenges that hinder the adoption and implementation of sound digital forensics schemes. Blockchain technology has been recently proposed as a viable solution for developing robust digital forensics mechanisms. In this paper, we provide an overview and classification of the available blockchain-based digital forensic tools, and we further describe their main features. We also offer a thorough analysis of the various benefits and challenges of the symbiotic relationship between blockchain technology and the current digital forensics approaches, as proposed in the available literature. Based on the findings, we identify various research gaps, and we suggest future research directions that are expected to be of significant value both for academics and practitioners in the field of digital forensics.

\end{abstract}

\keywords{Forensics, survey, review, blockchain, digital forensics}

\section{Introduction}

The undergoing digitization of information-intensive processes has a radical impact on our daily lives. Digitization affects almost all aspects of our lives from how we work to how we interact and communicate with each other. As a result, a myriad of devices is involved in almost every possible aspect of our daily lives. For instance, a smartphone may contain data with different levels of sensitivity (text messages, emails, financial transactions, etc.) which provide background information on its owner and his/her social connections. Digitization, however, comes at a cost. Financial frauds, intellectual property infringements, industrial espionage and digital terrorist networks are just a few among the various faces of cybercriminal behaviour. In the event of such deviant behaviour, digital evidence may be the only evidence of a case. Hence, digital evidence forms an integral part of the overall criminal investigation process.

Although the digital forensics community has established reliable scientific methodologies and common standards in its workflows, it still faces many challenges due to the volatile and malleable nature of the evidence and the continuous advances in technology that introduce new attack vectors. Moreover, most of the criminal activity on the Internet is transnational, generating cross-jurisdiction problems of cooperation and information exchange that can be alleviated by common standards and protocols. Several challenges have been identified in the literature regarding the development of robust digital forensics approaches. In \cite{Karie2015}, the authors identify four broad categories of challenges in digital forensics: a) technical challenges b) legal systems and/or law enforcement challenges c) personnel-related challenges and d) operational challenges. Another major issue that directly impacts digital forensics is the ever-increasing volume of potential evidence generated along with the growing number of devices used \cite{Quick2016}. Some domain-specific challenges are worth mentioning. For example, in the case of cloud forensics, identifying useful network events and recording the minimum representative attributes for each event remains a significant challenge \cite{Pilli2010}. Lack of international collaboration and legislative frameworks in cross-nation data access/exchange and the increased number of mobile devices accessing the cloud are also significant challenges in cloud forensics \cite{Ruan2013}. Arguably the most critical problem in digital forensics is the validity and trustworthiness of the evidence itself (safeguarding the chain of custody for the data related to a case), particularly when multiple stakeholders are involved in the overall forensics process.  

\subsection{Blockchain as a game-changer in digital forensics}

Blockchain, a novel disruptive technology, has emerged the past few years, enabling the development of a wide range of applications \cite{casino2018systematic}. In principle, a blockchain can be considered a distributed append-only data structure which stores states efficiently and in a transparent way. While the initial concept of Nakamoto was to store transactions of bitcoins in a way that prevents double-spending \cite{nakamoto2008bitcoin}, the created structure has many appealing properties. Setting aside the different ``flavours'' that blockchains have,  they offer auditability, robustness, and security. Blockchains also provide immutability to a large extent \cite{politoublock}, posing significant challenges to the implementation of the right to be forgotten principle, as defined in the EU General Data Protection Regulation Directive (GDPR) \cite{politou2018forgetting}.

Based on the above, it is apparent that the blockchain properties constitute a very promising baseline for forensics \cite{Al-Khateeb2019149}. More precisely, during a forensic investigation, all the involved investigators would like to store their findings in an immutable way so that they cannot be altered and be brought to a court of law. Similarly, blockchains provide transparency and auditability, which is a requirement for the chain of custody of the corresponding evidence. In this regard, in the past few years, several researchers have investigated these opportunities and proposed blockchain-based solutions for forensics. While the field is rather recent, we argue that in the coming years blockchain solutions for managing forensics will become a default. This survey performs an in-depth analysis of the needs and gaps of the field and the different approaches in the literature. Therefore, we set the landscape and facilitate the design of the new solutions.
 
\subsection{Goal and plan of the chapter}
In this chapter, we analyse the current state of blockchain-based forensic methods applied in different fields. First, we provide a comprehensive classification and the main features of the state-of-the-art solutions, which are retrieved using a sound bibliographic analysis approach. Next, we analyse how blockchain's features can enhance digital forensics. Finally, we discuss the limitations and the main challenges that are at the intersection of both fields.  

The rest of the chapter is organised as follows: Section \ref{sec:methods} describes the research methodology used and the main quantitative literature findings. Next, Section \ref{sec:classification} provides a topic classification of the blockchain-based digital forensics methods and a qualitative analysis of their features. Thereafter, Section \ref{sec:discussion} provides a discussion of the main limitations of blockchain technologies and the challenges to be faced by next-generation digital forensics solutions. Finally, the article offers some final remarks in Section \ref{sec:conclusions}.


\section{Methodology}
\label{sec:methods}

To survey the available blockchain-based forensics approaches, we have used a sound methodological framework. In particular, we performed a systematic search during May 2020 without time-frame restrictions. We used the Scopus scientific database as our primary source for identifying relevant literature. We used a predefined set of keywords for searching within the titles of all the available Scopus papers (the terms used included the words ``blockchain'' and ``forensics''). To locate additional studies, we used the so-called snowball effect (additional literature was retrieved based on references of key articles found in the initial phase of our search). We excluded some papers based on certain exclusion criteria (relevant to document type, structural quality, language, and subject area). In total, 24 articles were selected for analysis. For the thematic content analysis of the selected literature, we used a qualitative analysis software (MAXQDA11). Finally, we adopted various qualitative analysis methods (i.e., narrative synthesis and thematic analysis) for the classification and synthesis of the extracted data. We present the results of our analysis in Section \ref{sec:classification}.

\begin{table}[]
\centering
\begin{tabular}{|l|c|c|c|}
\hline
\multirow{2}{*}{\textbf{Publication type}} & \multicolumn{3}{c|}{\textbf{Publication year}} \\ \cline{2-4} 
 & \textbf{2018} & \textbf{2019}  &  \textbf{2020}\\ \hline
Journal articles & 2 & 10 &\\ \hline
Serials &  & 4 &\\ \hline
Conference proceedings & 4 & 7 & 1 \\ \hline
\end{tabular}
\caption{Year-based and source-type classification of the available literature}
\label{tbl:pubs}
\end{table}

There exist some bibliographic analysis results worth mentioning. As seen in Table \ref{tbl:pubs}, the available blockchain-enabled forensics literature spans only two years (2018 and 2019) with the exception, at the time of writing this paper, of one publication in 2020. Therefore, it is not until very recently that the scientific community has focused on blockchain technology as a viable solution for establishing robust forensics mechanisms. Regarding the type of publications, there seems to be an even allocation between conference proceeding papers and articles published in international journals.

\section{Classification of the available blockchain-based forensics literature}
\label{sec:classification}
\begin{figure}[th]
    \centering
    \includegraphics[width=\columnwidth]{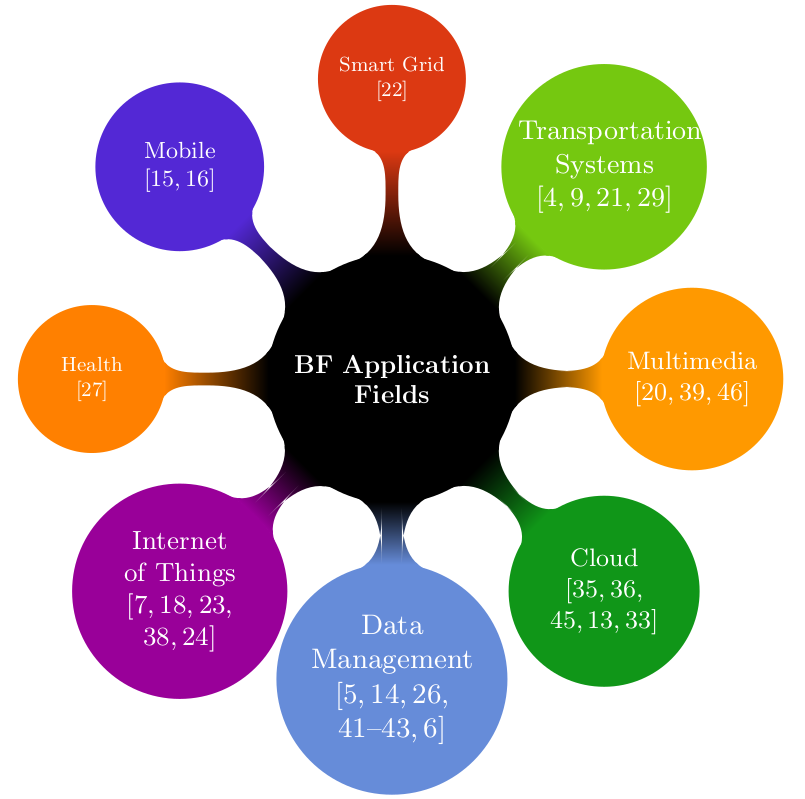}
    \caption{Mindmap abstraction the blockchain forensics research topics. The size of each topic has been weighted according to the number of contributions.}
    \label{fig:tikz}
\end{figure}

In this section, we thoroughly analyse the literature and provide a topic-based classification (see Figure \ref{fig:tikz}). Next, we identify the main features and solutions proposed by each method in Table 2 and discuss them in the following paragraphs.

\subsection{Cloud forensics}

Cloud security threats remain a significant challenge in nowadays. Cloud forensics, an umbrella term covering issues of cloud computing and digital forensics, may assist in investigating cloud environments and quickly respond to and report cloud security incidents \cite{Ricci2019}. Cloud forensics call for multiparty collaboration due to the multitude of stakeholders engaged. To this end, blockchain technology may enhance the collection of digital evidence in cloud environments and further improve different stakeholders coordination \cite{Zhang2018}. In this context of collaboration, security is mandatory, as recalled in \cite{Pourvahab2019153349}, where authors present a digital forensic architecture using fast-growing Software-Defined Networking (SDN) with robust data protection and access control. Another critical aspect of cloud forensics may refer to logs management. In particular, secure preservation and investigation of the various logs are essential elements of cloud forensics. However, due to the inherent uncertainties of cloud environment, several difficulties exist concerning the collection of authentic logs from a cloud environment while preserving integrity and confidentiality. Blockchain technology may be used as a logging-as-a-service tool for securely storing and processing logs while coping with issues of multi-stakeholder collusion and the integrity and confidentiality of logs \cite{Rane2019,Duy2019416}.

\subsection{Data management forensics}
The works classified in data management include these proposing novel models for data processing and chain of custody preservation methodologies. The use of permissioned blockchains \cite{Gopalan2019,Lone2019,Xiong2019} is stated as a measure to enhance scalability, as well as the use of lightweight consensus mechanisms \cite{Tian2019}. Advanced evidence collection and feature classification \cite{Billard2018}, as well as the relevance of the timeline of events and chain of custody \cite{Weilbach2019,Bonomi2020} are other features discussed by authors. However, the main drawback of the proposed solutions is that they only offer architectural designs and they do not provide full exploitation of blockchain, with only a few of them offering practical implementations \cite{Lone2019,Tian2019,Bonomi2020}.

\subsection{Healthcare forensics}

With the prevalence of new regulatory frameworks brought forward (like the EU GDPR directive), health care organizations have started taking necessary steps towards protecting themselves against costly breaches of patients' sensitive information and further safeguarding their reputation. To this end, forensics may be a valuable ally for addressing litigation risks when it comes to data breaches and unauthorized access to medical data from both outside attacks or internal misuse \cite{Chernyshev2019}. Access control management is an essential feature of patient data protection. In \cite{malamas2019forensics}, the authors propose a blockchain-enabled authorization framework for managing both the Internet of Medical Things (IoMT) devices and health care stakeholders. The proposed framework provides fine-grained access to patient health data and preserves the chain of custody of all logs by offering audit trails for integrity and provenance guarantees.

\subsection{IoT forensics}

IoT forensics includes the study of IoT devices, their systems and interrelations between different parts of their ecosystems. In this regard, the result of our literature review showed that there is a relevant interest in IoT forensics in the blockchain. We observed that evidence collected from IoT devices and interactions between the different actors (e.g. through privacy-preserving identity management techniques) are the most relevant features studied in the literature\cite{Brotsis2019,Hossain2018,Ryu2019}. Moreover, proper identity management and privacy preservation is also a mandatory requirement in such context \cite{Le2019}. In the case of \cite{Li20191433}, authors propose a framework for evidence storage and examination based on structured data logs, guaranteeing the provenance and auditability of each evidence item. Nevertheless, current solutions are not mature enough, since authors only proposed architectures and flows, except for \cite{Le2019}, which only provided transaction performance tests.

\subsection{Mobile forensics}
Mobile forensics includes the analysis of digital and physical evidence provided by smartphone devices and similar ones (i.e. these sharing similar architectural bodies and underlying operating systems, such as tablets or other handheld devices). Nevertheless, the identified blockchain-based forensic research mainly focuses on applications and malware detection. More concretely, authors propose the use of consortium blockchains and focus on malware detection and statistical analysis based on each application feature \cite{Gu2018,Homayoun2019}. Therefore, more work needs to be provided in this field, with special regard to hardware inclusion and holistic systems definition, as well as usable implementations.

\subsection{Multimedia forensics}

Multimedia Forensics employs various scientific techniques for examining a multimedia file (audio, video and/or image) concerning its a) integrity (establish the linkage between a multimedia output and its source identification) and b) authenticity (check for the veracity of the multimedia output). For example, in \cite{Kerr2019} a blockchain-based approach is proposed for cataloguing CCTV video evidence. The authors provide a functional implementation of the blockchain-based system that manages high volumes of CCTV evidence. In \cite{Samanta2018} the authors present E-Witness, a system that uses blockchain technology for safeguarding the integrity and spatio-temporal characteristics of digital evidence captured by smartphones. To verify the integrity and spatio-temporal claims of the evidence, the proposed system uses hashes of pictures/videos along with location certificates stored in the blockchain. A blockchain-based photo forensics scheme is presented in \cite{Zou2019}. The proposed Ethereum-based scheme resolves photos' veracity issues like photo-faking, photo-tracing and copyright dispute problems.

\subsection{Smart grid forensics}

Smart grids offer significant improvements in terms of resources utilization in current electricity supply networks. Smart grids embrace digital communications technologies, smart metering, intelligent appliances and energy-efficient resources for better matching energy supply and demand. Like other cyberphysical systems, however, smart grids are vulnerable to cyberattacks, and intrusion detection might prove extremely important. In \cite{Kotsiuba2019} the theoretical underpinnings of blockchain technology and its importance in smart grids forensics are discussed. The authors highlight how blockchain can enhance features such as energy optimisation, system performance, managerial tasks, and security of smart grids. Finally, the authors discuss the opportunities/open issues in the topic.

\subsection{Intelligent Transportation Systems forensics}

Intelligent Transport Systems (ITS) embrace a range of technological novelties like advanced sensing and control, and IoT applications for improving safety, efficiency and services provision of both vehicles and road transport networks. However, the increased automation of ITS (e.g. self-driving cars) and the adoption of new data privacy frameworks (like the GDPR) call for the development of sound forensic mechanisms to analyze traffic accidents and protecting users' sensitive data. In \cite{Billard2019} a blockchain-enabled system is proposed for managing users requests (car navigation) and relevant data that fully complies with data privacy and protection legal frameworks. In \cite{Cebe2018} the authors propose a blockchain-based forensics system that enables the trustless, traceable, and privacy-aware post-accident analysis with minimal requirements in storage and processing. A blockchain framework is proposed in \cite{Kevin2019} for managing sensitive navigation data (GPS position) within a fixed geographic zone while ensuring user anonymity. Cybersecurity threats may also prove critical in the context of current ITS. In \cite{Patsakis2019} a blockchain-based framework is proposed for keeping logs of all hardware profile changes in a vehicle. Based on the inherent characteristics of blockchain technology, the proposed framework only allows authenticated changes, subject to user, time, geospatial, and contextual constraints, as defined by automotive manufacturers. 

\begin{landscape}
\setcaptiontype{table}
\rowcolors{2}{gray!25}{white}
\scriptsize
\centering
\caption{Description of the features of the available blockchain-based forensic schemes.}
\begin{longtable}{cp{1.1in}p{2.2in}p{2.7in}c}
\toprule   
\textbf{Ref.} & \multicolumn{1}{c}{\textbf{Application domain}}  & \multicolumn{1}{c}{\textbf{Problem addressed}} & \multicolumn{1}{c}{\textbf{Blockchain-enabled forensic features}} & \multicolumn{1}{c}{\textbf{Implementation}} \\
\midrule 
\endhead
\cite{Rane2019} & Cloud forensics & integrity of logs, multi-stakeholder collusion &secure logging as-a-service for cloud environment, integrity, confidentiality and immutability of logs & Yes \\
\cite{Ricci2019} & Cloud forensics & multi-location storage of forensic evidence, multiple stakeholders engaged & data encryption, distributed storage & No \\
\cite{Zhang2018} & Cloud forensics & multiparty cooperation, trustworthiness of records among stakeholders & chain of custody, proof of existence, privacy and anti-tampering preservation for process records & Yes \\
\cite{Pourvahab2019153349} & Cloud forensics & multi-stakeholder collusion, security and access control,  multiparty cooperation, trustworthiness of records among stakeholders &  integrity, chain of custody, data encryption, secure access control  & Yes \\
\cite{Duy2019416} & Cloud/network forensics & multiparty cooperation, trustworthiness of records among stakeholders, SDN log recording & chain of custody, proof of existence, privacy and anti-tampering preservation for process records & Yes \\
\cite{Billard2018} & Data management forensics & validity of the digital evidence & weighted digital evidence,  digital  evidence inventory, categorization according to each evidence relevance, assignment of confidence rating & No \\
\cite{Bonomi2020} & Data management forensics & integrity and validity of electronic evidence and ownership & chain of custody, tracking of the stakeholders involved, credibility of the data provided & Yes \\
\cite{Gopalan2019} & Data management forensics & integrity and validity of electronic evidence & chain of custody, tracking of the stakeholders involved, credibility of the data provided& No \\
\cite{Lone2019} & Data management forensics & Integrity and authenticity of digital evidence, authenticity and legality of processes and procedures used to gather and transfer the evidence & chain of custody, safeguarding the integrity  and tamper-resistance of digital forensics & Yes \\
\cite{Tian2019}& Data management forensics & tampering with evidence, data privacy issues, sensitive information leakages & lightweight, scalable secure digital evidence framework, multi-signature schemes for evidence submission and retrieval & Yes \\
\cite{Weilbach2019} & Data management forensics & Proof of existence of digital evidence & tamper-proof chronology by means of OpenTimestamps& No \\
\cite{Xiong2019} & Data management forensics & trust and security issues as derived by current centralized data management schemes & electronic evidence preservation, different evidence access rights, data security  protection, information integrity guarantees, traceability & No \\
\cite{malamas2019forensics} & Health care forensics & different access levels (to both health data and  devices), health data privacy & log audit trails for integrity and provenance guarantees, health data privacy, fine-grain access & Yes \\
\cite{Brotsis2019}& IoT forensics & collection and preservation of evidence regarding alleged malicious behavior in IoT networks & private forensic data/metadata evidence collection, integrity, authentication,  and non-repudiation of the data collected & No \\
\cite{Hossain2018}& IoT forensics & multiple IoT stakeholders, multi-party access to  digital evidence & integrity, confidentiality, anonymity, authenticity and non-repudiation & No \\
\cite{Le2019} & IoT forensics & traceability, integrity and provenance of the evidence is limited due to the resource-constraint nature of IoT devices & integrity, authenticity, non-repudiation,  identity privacy, end-to-end forensic life cycle & No \\
\cite{Ryu2019}& IoT forensics & heterogeneity and distribution characteristics of the  IoT environment & chain of custody for all the IoT-related forensics processes, security and data integrity, multi-party verification of the IoT-related forensics processes & Partial \\
\cite{Li20191433} & IoT forensics & traceability, integrity and provenance of the evidence & security and data integrity, multi-party verification of the IoT-related forensics processes and evidences & Partial \\
\cite{Gu2018}& Mobile forensics & Current limitations  of static-based and dynamic-based code analysis tools (code obfuscation, encryption, malware in different families with various features) & consortium blockchain framework to store and classify android malware, classification of different malware families & Partial \\
\cite{Homayoun2019}& Mobile forensics & tracking and recording of a very wide range of existing malicious programs, current limitations  of static-based  and dynamic-based code analysis tools & enhanced malware detection features based on the usage of both private and consortium blockchain & No \\
\cite{Kerr2019} & Multimedia forensics & integrity and legal authenticity of video data produced as evidence in legal proceedings, privacy concerns of  video data gathered by CCTV installations & trustworthy evidence protection in distributed network environment, video data integrity (link to the primary video stream and its creation & Yes \\
\cite{Samanta2018} & Multimedia forensics & Civilians/journalists who need to protect their identity while ensuring that the evidence they collect are forensically sound& integrity and spatio-temporal properties of digital evidence & Yes \\
\cite{Zou2019}& Multimedia forensics & photo-faking, photo owners have limited control over their photos after uploading them on the Internet due to lack of copyright protection mechanisms & customized access control rules,  photo-tracing, creation of copyright-protected photos (resolving copyright dispute problems)& Yes \\
\cite{Kotsiuba2019}& Smart grids forensics & smart grid security, intrusion detection & ensure the integrity of smart energy transaction  platforms, keeping log information for effectively investigate cybercrimes and predict system failures & No \\
\cite{Billard2019} & Transportation forensics & contradictory use of personal data, privacy, multiple stakeholders involved& integrity, veracity, authenticity, non-repudiation  and identity privacy of vehicle-related data,  voluntarily and spontaneous release of data for forensic purposes & No \\
\cite{Cebe2018} & Transportation forensics & transportation data is overwritten shortly, no available system for integrating data from the various stakeholders involved (data from other vehicles, road conditions, manufacturers, and maintenance centers), only third party solutions exist for vehicular forensics (such as surveillance cameras and eyewitnesses). & lightweight privacy-aware blockchain framework  to manage the collected vehicle-related data  (maintenance information/history, car diagnosis  reports) & No \\
\cite{Kevin2019} & Transportation forensics & data privacy concerns (GPS sensitive info) due to third  parties usage & legal authority may run forensic analysis without unnecessary violation of the user anonymity and privacy & No \\
\cite{Patsakis2019} & Transportation forensics & unauthorized changes in Vehicle Hardware Profiles, multiple stakeholders involved & logs of all hardware profile changes are kept on blockchain, provision of customized access (only authenticated changes are allowed) & Yes \\\bottomrule
\end{longtable}
\end{landscape}

\section{Discussion}
\label{sec:discussion}
In what follows, we describe the main limitations of blockchain technology and some strategies to overcome them. Moreover, we provide a detailed analysis of actual and future challenges of digital forensics and discuss possible countermeasures.

\subsection{Limitations in blockchain}


The suitability of blockchain is a topic that has been extensively discussed in the literature \cite{casino2018systematic,8902067}. In this regard, the challenges to be faced by different blockchain technologies vary depending on their type and application scenario. For example, public blockchains face limitations such as scalability, performance, and cost issues. In this regard, public blockchains are nowadays mainly used for cryptocurrencies and to commit small pieces of data (i.e. hashes) for verifiability purposes \cite{yousaf2019tracing}. In the case of private blockchains, the performance and scalability challenges are overcome due to the use of more efficient consensus mechanisms and a reduced number of participants.

Moreover, the cost of memory, compared with public blockchain is negligible, yet off-chain data storage is a recommended strategy for most applications. Nevertheless, both public and private blockchains require the use of proper data management and architectural designs to provide security and privacy guarantees \cite{li2017survey}. In this regard, the use of secure identity management systems \cite{8888155}, the proper analysis of the specific blockchain systems to be used \cite{homoliak2019security}, and a careful implementation development of smart contracts \cite{SINGH2020101654,atzei2017survey} are mandatory.

\subsection{Challenges in Blockchain Digital Forensics}
We classified next-generation digital forensics most relevant challenges in the following six domains:

\subsubsection{Tokenization of artifacts from digital evidence}
Digital forensics imply the analysis of the digital evidence and the extraction of the corresponding knowledge regarding the events of a crime under investigation. However, this analysis is not performed by a single entity. For instance, an image of hard disk may contain different evidence that must be analysed by different people who will look into different parts. One person may study the log files, while another may investigate the file system and a third one might be needed to analyse a specific binary that requires reversing. Therefore, a single evidence is expected to be divided in an arbitrary amount of artifacts, each of which might have to be studied individually and from another person. Breaking down things and storing them in blockchains is not straightforward and several existing solutions could be adopted (e.g. the use of tokens), however, the bulk of them considers that the elements that something is decomposed to is predetermined. Despite the fact that some solutions for assigning tokens in blockchain for arbitrary decomposition of an object have been proposed in the supply chain field \cite{dasaklistokens}, storing tokenised artifacts in the blockchain during the course of a digital investigation remains a challenge.

\subsubsection{Efficient management of data volume in the chain of custody.}
One of the main concerns in digital forensics is the volume of data, since evidence may include thousands of multimedia files or log files per case. In this regard, although data storage of raw documents has to be provided for all cases, it should be based on off-chain technologies (e.g. IPFS, Storj). In this case, only hashes should be used in the blockchain (i.e. or meta-hashes if data are processed as blocks, to ease auditability). 

\subsubsection{Parse forensic sound procedures in blockchain systems.}
Standard and sound forensic flows have to be provided, even when using blockchain as a platform to provide verifiability and chain of custody tamper-proof guarantees. Therefore, proper standardized flows and smart contracts that map the adequate functions have to be provided to enable final court validation as well as certification by digital forensic laboratories and law enforcement agencies. 

\subsubsection{Enable an understandable forensic outcome/reports.}
The use of blockchain provides a myriad of benefits, such as the efficient and verifiable provision of data flows. Still, the knowledge retrieving and report creation parts belong to a different stage. In this regard, even if automated, the reports and outcomes generated should be understandable in court. Therefore, even if blockchain facilitates this task, research efforts have to be done in this direction, providing a link between forensic sound procedures and their proper explanation.

\subsubsection{Interoperability and cross-border jurisdictions.}
The use of international standardized flows and proper data management and sharing agreements will enhance the fight of cybercrime. Nowadays, international collaborations already exist in the scope of the European Union\footnote{https://ec.europa.eu/home-affairs/what-we-do/policies/organized-crime-and-human-trafficking/e-evidence\_en}. Nevertheless, further development of blockchain-based solutions\footnote{https://locard.eu/} will serve as a ground truth platform for standardised solutions, enabling international interoperability.

\subsubsection{Timeline of events and chronology.}
The relevance of data acquisition and timeline of events in digital forensics is key to identify patterns and relate similar cases, since the knowledge generated by forensic investigations has to be used in the future to prevent or minimise them. Therefore, the proper reporting and evidence collection procedures have to be done respecting the timeline of events. Blockchain can provide proof-of-existence due to its immutability, which combined with the use of block timestamps and hashes, can guarantee that evidence was collected at a specific moment and they have not been modified.


\section{Conclusions}
\label{sec:conclusions}
Digitization comes with a myriad of novel opportunities and services. Nevertheless, this heterogeneous landscape is also becoming a profitable playground for malicious users, which are continuously increasing the dynamism and complexity of cybercriminal activities. In this regard, digital forensics needs to be rapidly updated to deal with a set of multidisciplinary challenges, ranging from the advances in information and communication technologies, to jurisdictional and interoperability restrictions. To this end, we believe that digital forensics can be benefited from the widespread adoption of blockchain technology and its inherent characteristics.

In this chapter, we presented a literature review of the current blockchain-based forensic solutions and classified them according to their features as well as their application field. Thereafter, we identified the benefits and limitations of blockchain-based forensics and outlined the main challenges to be overcome in the future, providing a fertile ground for research.

Future work will focus on developing a blockchain-based forensic framework which enables the collection of heterogeneous digital evidence as well as forensic procedures in an standardised manner. To this end, we will study the tokenisation of digital forensic evidences to provide a common layer of abstraction for different categories of cybercrime.  

\section*{Acknowledgments}
This work was supported by the European Commission under the Horizon 2020 Programme (H2020), as part of the project  \textit{LOCARD} (\url{https://locard.eu}) (Grant Agreement no. 832735).

\bibliographystyle{splncs}
\bibliography{bibliography.bib}

\end{document}